\documentclass[superscriptaddress,twocolumn,amsmath,amssymb]{revtex4}
\usepackage{graphicx}% Include Figure files
\usepackage{dcolumn}% Align table columns on decimal point
\usepackage{bm}% bold math
\usepackage{latexsym}
\begin{document}
%%%%%%%%%%%%%%%%%%%%%%%%%%%%%%%%%%%%%
\title{Centenary of ``Researches on irritability of plants'' by Jagadis 
Chandra Bose}
\author{Debashish Chowdhury{\footnote{J.C. Bose National Fellow}}}
%\email{debch@iitk.ac.in}
\affiliation{Department of Physics, Indian Institute of Technology,
        Kanpur 208016, India}
\date{\today}
%%%%%%%%%%%%%%%%%%%%%%%%%%%%%%%%%%%%%%%
\begin{abstract}
This note celebrates the centenary of Jagadis Chandra Bose's classic monograph entitled 
``Researches on irritability of plants''. 
\end{abstract}
\maketitle
%%%%%%%%%%%%%%%%%%%%%%%%%%%%%%%%%%%%%

\begin{figure}[htbp]
\begin{center}
\includegraphics[width=0.95\columnwidth]{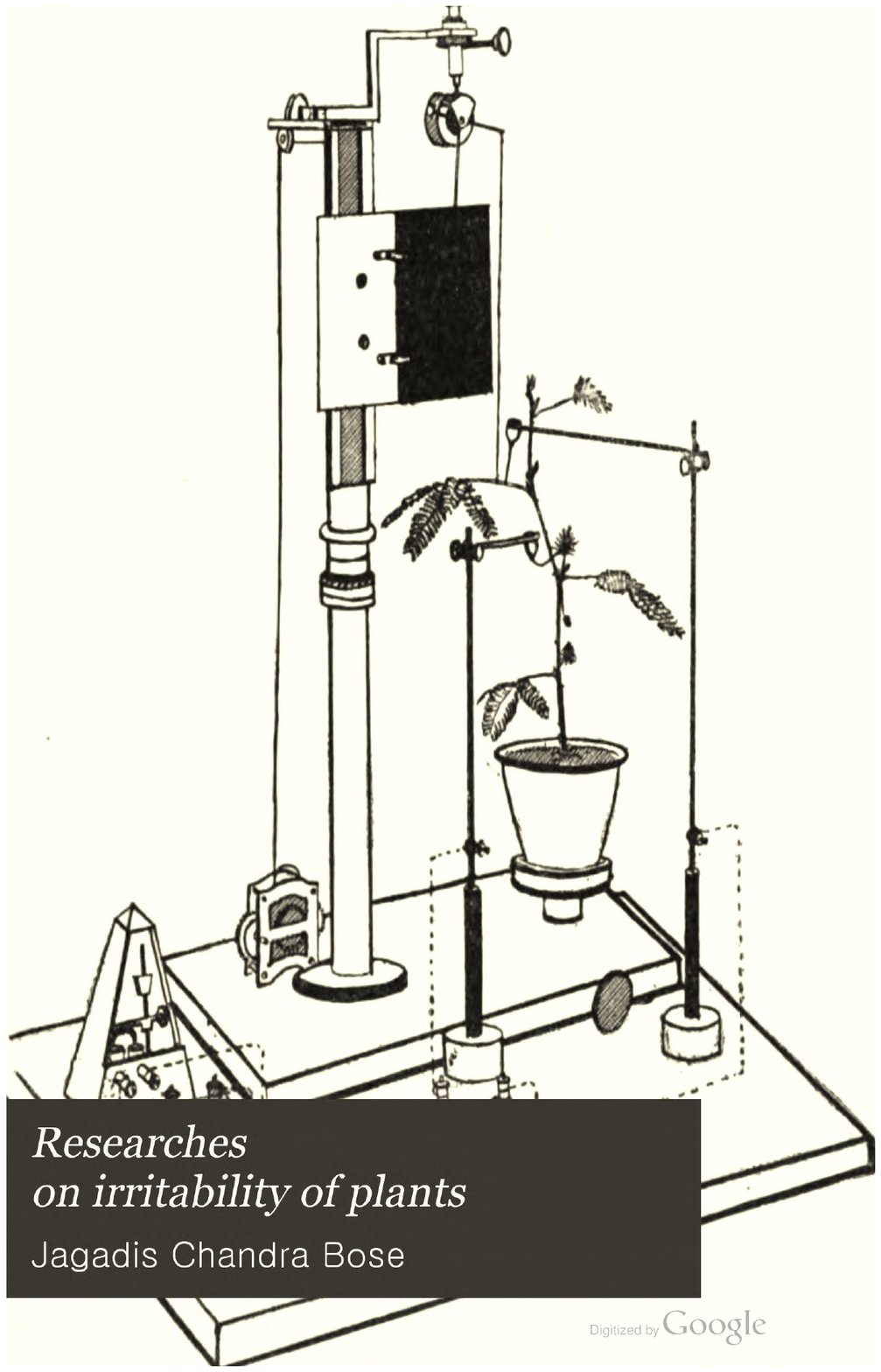}
\end{center}
\caption{The cover page of the monograph ``Researches on irritability of plants'' authored by 
Jagadis Chandra Bose and published by Longmans, Green and Co. in 1913.
}
\label{fig-coverpage}
\end{figure}

In 1913 Sir Jagadis Chandra Bose published a monograph (see the coverpage in fig.\ref{fig-coverpage}) 
``Researches on irritability of plants''' \cite{bose1913}. 
The concept of {\it irritability} \cite{verworn1913} had its origin in the works of Francis Glisson and  Albrecht von Haller \cite{haller1801}. On the basis of experiments performed on animals, Haller emphasised the distinction between `sensibility' and `irritability'- the ability to perceive a stimulus is sensibility whereas the ability to respond to that stimulus is called irritability. For example, nerves are the vehicles of sensibility whereas muscle contraction is one mode of irritability. 
Till the end of the nineteenth century, the experimental investigations of irritability were carried out, almost exclusively, with animals. Sir J.C. Bose was one the few visionaries who did the pioneering works on the irritability of plants with a very sophisticated equipment that he himself developed for this purpose. In this commemorative article I briefly discuss the aims of the investigations reported in this classic and the impact of the results. I also mention how irritability depends on the operation of molecular machines, a concept pioneered by Marcello Malpghi \cite{piccolino00} in the sixteenth century and a subject of current mutidisciplinary research in the twentieth century \cite{chowdhury13a,chowdhury13b}. 

%%%%%%%%%%%%%%%%%%%%%%%%%%%%%%%%%%%%%%%%%%%%%%%%%%%%%%%%%%%%%%
%%%%%%%%%%%%%%%%%%%%%%%%%%%%%%%%%%%%%%%%%%%%%%%%%%%%%%%%%%%%%

Bose began his monograph with the opening observation \cite{bose1913}: ``In strong contrast to the energetic animal, with its various reflex movements and pulsating organs, stands the plant in its apparent placidity and immobility. Yet the same environment, which with its changing influences so strikingly affects the animal, is playing upon it also''. He reminded the reader \cite{bose1913}: ``Storm and sunshine, the warmth of summer and the frost of winter, drought and rain, all these and many more come and go about it''. Then he pointedly asks the main question \cite{bose1913}: ``What coercion do they exercise upon it? What subtle impress do they leave behind?'' 

In the same introductory chapter Bose makes clear concrete statements on the quantities to be mesaured \cite{bose1913}: ``In studying the excitatory reactions of the plant, under external stimulus, we have to determine, first, what time elapses between the incidents of the shock and the initiation of a perceptive responsive movement. This constitutes the determination of the {\it Latent Period}. We have next to find out at what rate this responsive movements of the leaf takes place, and after what time the contractile phase of the movement is exhausted. ...We therefore want to know the various rates at which recovery gradually takes place''. Summarizing the observations made with his own equipment, Bose concluded \cite{bose1913}: ``In surveying the response of living tissues we find that there is hardly any phenomenon of irritability observed in the animal which is not also found in the plant.'' 

Burton E. Livingston reviewed this monograph in {\it Science} \cite{livingston1914}. He praised Bose's apparatus writing that it ``is so ingenious and delicately efficient as to excite wonder and admiration in and for itself'' \cite{livingston1914}.  He recommended the book to research workers who, he wrote, ``will find this book replete with novel ideas and novel ways of attaining quantitatively comparable measures of plant irritability'' \cite{livingston1914}. 

William Crocker reviewed this monograph in {\it Botanical Gazette} \cite{crocker1914}. He praised the ``excellence of methods and execution'' of Bose's experiments. He also noted the ``direct, clear'' presentation as well as the ``comprehensive but concise summaries at the end of each chapter and at the end of the volume'' \cite{crocker1914}. Crocker nicely highlighted the implications of Bose's observations in plant physiology \cite{crocker1914}: ``in some cases confirming views already held; in others showing prevailing views at error; and in still others giving exact determination of physiological critical time periods''. However, it was regrettable that this ``excellent work'' was marred at a few places  by the ``oldness of viewpoint and lack of knowledge of physiological literature'', particularly ``modern knowledge of the physics and chemistry of living cells and plant response'' \cite{crocker1914}. 

As chronicled in his biography by Patrick Geddes \cite{geddes1920}, Bose proceeded on a scientific deputation in 1914 at the invitation of several European and American academic institutions where he lectured on irritability of plants. His lecture session at Cambridge was presided over by Francis Darwin, one of the leading contemporary botanists who collaborated with his father Charles Darwin, on pioneering research on the movement of plants \cite{darwins}. 

How is irritability, particularly, that of plants, explained in the light of our current knowledge on the molecular structure and dynamics of living matter? In the 18th century, in the absence of any structural information at the microscopic level, one could only speculate that, perhaps, irritability arises from some internal organization.  Nevertheless, concept of irritability laid the foundation for Claude Bernard's 19th century concept of `internal milieu' \cite{barcroft1931} and Walter B. Cannon's 20th century concept of `homeostasis' \cite{cannon1932}.  According to the latter, the body not only maintains a non-equilibrium steady state, but also restores this state when perturbed by external causes. Some physiologists, including Bernard, revived an earlier concept of machines, which are essentially working components of the body, to explain this phenomenon. But, what was the original concept of machines in living body?

In the 17th century Marcello Malpighi \cite{piccolino00} speculated the existence of tiny machines, invisible to the naked eye, in the organs.  However, because of both technical hurdles and the emergence of new attractive areas of investigation, research on molecular machines was practically abandoned. Interestingly, for the potential readers Livingston \cite{livingston1914} clarified that Bose's work \cite{bose1913} was concerned with the irritability of the entire plant or with that of its organs; his monograph did not deal with irritability of plant cells. However, since cell is the structural and functional unit of life, its irritability is of fundamental importance in modern molecular cell biology. 

Interest on molecular machines was revived in the 20th century when structural studies of these became feasible. Individual molecular machines could be ``caught in the act'' only towards the end of the 20th century. In a very influential article \cite{alberts98}, written towards the end of the twentieth century, Bruce Alberts, then president of the US National Academy of Sciences, inspired the new generation of cell biologists to view a living cell as a micro-factory. Coordinated operation of large number of nano-machines drives the key intracellular processes in a cell \cite{chowdhury13a}. These machines can be broadly divided into two groups: (i) machines involved in intracellular motility and contractility, (ii) machines involved in the genomic processes.  

Movement, the most obvious mode of irritability of living matter, is also regarded as the hallmark of life \cite{chong00}. The cause of many different forms of movements in animals can be traced to the machine-driven motility and contractility at the cellular sub cellular levels \cite{chowdhury13a}. In animals, contraction of muscle, for example, is caused by a relative sliding of two filaments driven by a superfamily of motor proteins called {\it myosin} that are powered by the hydrolysis of adenosine triphosphate (ATP) to adenosine diphosphate (ADP) \cite{chowdhury13a}.  But, plants don't have muscles although a muscle-like sliding filament mechanism explains the phenomenon of cytoplasmic streaming in plant cells \cite{shimmen07}. Movement of plants \cite{koller11}, which has received attention of pioneers like Darwins \cite{darwins}, are driven dominantly by mechanisms quite different from those in animals \cite{leopold00}. Plants exploit water \cite{dumais12} for controlling large internal (turgor) pressure in cells which, in turn, is utilised for its own movements that span a wide range of time scales. But, the machineries for driving the genomic processes in animals and plants display lot more similarities. 
Response of plant and animal cells to external environmental stress is an active area of current research \cite{nadal11}. Understanding how the plants respond to environmental stress can help us in efficient crop management to overcome the potential threats to food security posed by the global climate change \cite{mcclung10}. 

Bose's curiosity in the plant response to external stimuli was a natural continuation of his earlier interest, as a physicist, in the nature of response of non-living systems to external forces which could be of mechanical or electro-magnetic origin. In his own words (in the preface of ref.\cite{bose1926} ``My investigations on plant-response date from the discovery of electric response of non-living matter, such as metals''. 
The monograph of 1913  was neither Bose's first nor the last work on plant response. The monograph of 1913 was preceded by two of his monographs on related problems \cite{bose1902,bose1906,bose1907}. But, his monograph of 1926, which he dedicated to his `life-long friend Rabindra Nath Tagore', had the provocative title `The nervous mechanism of plants'. He was aware that ``..connecting nervous links had not been suspected in the plant''. But, he claimed that the  researches described in his monograph \cite{bose1926}  show that "not only has a nervous system been evolved in the plant, but that it has reached a very high degree of perfection, as marked by the reflex arc in which a sensory becomes transformed into a motor impulse.'' 

Finally, how was Bose's plant research received by his contemporary botanists and what was the impact of this work on research in plant physiology over the next 100 years? 
Bose was knighted in 1917 and elected a fellow of the Royal Society in 1920. But, not all western scientists were `Bosephiles'; there were many `Bosephobes' who were critical of his methods and skeptical of his claims \cite{shepherd12}. Was he ``intermixing non-reproducible data and startling claims with plausible descriptions'', as Barbara Pickard \cite{pickard73} wrote many years later? The reality is that his data could not be reproduced by the research groups of many of his contemporaries simply because they could not set up equipments of comparably high accuracy. Neverthless, Pickard \cite{pickard73} gave full credit to Bose for reporting some of his quantitative data which were found to be correct by later workers. Some authors, like Wayne,  believe that ``racism and sexism'' \cite{wayne94} might have been the non-scientific reasons why Bose's work and later work of Pickard , respectively, on electrical communication in plants were somewhat overlooked.  Some of the terms, such as plant nerves, that he used as metaphor might have been misinterpreted by his contemporaries. Plant scientists are still debating whether the use of such metaphors, including the more recent provocative term `plant neurobiology' \cite{brenner06}, are  `essential adjunct to the imaginative scientific mind'  \cite{trewavas07,brenner07,barlow08} or has ``limited scientific benefits'' \cite{alpi07}. The tree of knowledge on `nervous mechanism of plants' is growing- Bose sowed the seed of this plant.

My research has been supported by Dr. Jag Mohan Garg Chair professorship (IIT Kanpur), by J.C. Bose National Fellowship (SERB, DST, government of India), and by a research grant from DBT (government of India).

%%%%%%%%%%%%%%%%%%%%%%%%%%%%%%%%%%%%%%%%

\end{document}